\documentclass[12pt]{article}

\usepackage[utf8]{inputenc}
\usepackage[utf8]{inputenc}

\usepackage{amssymb}
\usepackage{amsfonts}
\usepackage{graphicx}
\usepackage{circuitikz}
\usetikzlibrary{quantikz}

\usepackage[verbose]{placeins} 

\usepackage{braket}
\usepackage{tikz}
\usepackage[T1]{fontenc}
\usepackage{imakeidx}

\usepackage{xcolor,varwidth}

\usepackage{pgfplots}
 \usetikzlibrary{backgrounds,calc}
 \usepackage{amsmath,tikz}
 \usetikzlibrary{quotes,angles}
\pgfplotsset{compat = newest}
\usetikzlibrary{quotes,angles}

\usetikzlibrary{svg.path}
\usepackage{pgfornament}

\usepackage{xcolor}

\usepackage{subfig}
\usepackage{graphicx}
\def\mset #1[#2]=#3{%
	\expandafter\xdef\csname #1#2\endcsname{#3}
}
\def\mget #1[#2]{%
	\csname #1#2\endcsname
}
\def\minc #1[#2]+=#3{%
	\pgfmathparse{\mget #1[#2]+#3}%
	\mset #1[#2]=\pgfmathresult
}

\date{}

\usepackage{hyperref}
\usepackage [all]{hypcap}
\usepackage{color}

\usepackage{float}

\usepackage{listings}
\definecolor{codegreen}{rgb}{0,0.6,0}
\definecolor{codegray}{rgb}{0.5,0.5,0.5}
\definecolor{codepurple}{rgb}{0.58,0,0.82}
\definecolor{backcolour}{rgb}{0.95,0.95,0.92}

\lstdefinestyle{mystyle}{
    backgroundcolor=\color{backcolour},   
    commentstyle=\color{codegreen},
    keywordstyle=\color{magenta},
    numberstyle=\tiny\color{codegray},
    stringstyle=\color{codepurple},
    basicstyle=\ttfamily\footnotesize,
    breakatwhitespace=false,         
    breaklines=true,                 
    captionpos=b,                    
    keepspaces=true,                 
    numbers=left,                    
    numbersep=5pt,                  
    showspaces=false,                
    showstringspaces=false,
    showtabs=false,                  
    tabsize=2
}

\begin{document}

\title{A novel  quantum circuit for the quantum Fourier transform  }
\author{ Juan M. Romero \thanks{jromero@cua.uam.mx },  Emiliano Montoya-Gonz\'alez \thanks{emiliano.montoya.g@cua.uam.mx },   
Guillermo Cruz,\\  \thanks{luis.cruz.e@cua.uam.mx} and Roberto C.  Romero \thanks{roberto.romero@cua.uam.mx }\\
Departamento de Matemáticas Aplicadas y Sistemas,\\
Universidad Aut\'onoma Metropolitana-Cuajimalpa,\\
M\'exico, D.F 05300, M\'exico }
\date{\today}

\maketitle
\begin{abstract}
The Quantum Fourier Transform (QFT) is a fundamental component of many quantum computing algorithms. In this paper, we present an alternative method for factoring this transformation. Inspired by this approach, we introduce a new quantum circuit for implementing the QFT. We show that this circuit is more efficient than the conventional design. Furthermore, using this circuit, we develop  alternative versions of the HHL algorithm and Shor's algorithm, which also demonstrate improved performance compared to their standard implementations.\\
 

\end{abstract}

\section{\label{sec:introduction}Introduction}

For certain algorithms,  quantum computing  has demonstrated better performance than classical computing.
For instance, in classical computing the Fast Fourier Transform  is one of the most important algorithms,  which has computational complexity of $O(n2^{n})$. However,  the Quantum  Fourier Transform (QFT)   has $O(n^{2})$  computational complexity \cite{Schuld}.  The  QFT  plays a fundamental role for different quantum algorithms. For example, the Shor’s factoring algorithm is one of most relevant  in quantum cryptography and the QFT is one of its building block  \cite{Shor}. Moreover, the Harrow-Hassidim-Lloyd algorithm (HHL) provides a quantum method  for solving linear systems of equations and  for this algorithm the QFT plays a fundamental role \cite{HHL}. It is worth  mentioning that the HHL algorithm has applications in quantum machine learning \cite{Conti,Alchieri},  physics \cite{he,nuclear}, quantum chemistry \cite{chemistry}, finance \cite{Herman}, etc.
Given that the QFT is a cornerstone of quantum computing, various alternative formulation have been proposed to enhance its performance, including, approximate  \cite{Approx},  fast   \cite{Asaka}, and semiclassical versions  \cite{Semi}.\\

In this paper, an alternative method for factoring the QFT is presented.  Inspired 
by this method, a new quantum circuit for the Quantum Fourier Transform is introduced. Moreover, it is shown that the new
algorithm is faster than the usual. Furthermore, using this quantum circuit,  alternative HHL algorithm and Shor's algorithm are presented, demonstrating better performance than their conventional implementations.\\

This paper is organized as follows.  In Sec. \ref{RQFT} a brief review of the QFT is provided.  
 In Sec. \ref{AQC} an alternative quantum circuit for the QFT is given.  In Sec. \ref{AHHL} an alternative quantum circuit for the HHL algorithm is presented. 
   In Sec. \ref{Shor} an modified  quantum circuit for the Shor's algorithm is presented. In Sec. \ref{Con} a summary is given. 

\section{Review of QFT }
\label{RQFT}

First, lets us remember that a basis of  a system with $n$ qubits have $2^{n}$ states, for example
 \begin{eqnarray}
 & &\underbrace{\ket{00\cdots 0 0}}_{n}, \label{base1}  \\  & &\underbrace{\ket{00\cdots 0 1}}_{n}, \label{base2} \\ & &\vdots  \nonumber \\&  &\underbrace{\ket{11\cdots 1 1} }_{n} \label{basen}.
 \end{eqnarray}     
In addition, notice that by using the binary numeral system the first $2^{n}$  natural numbers
 \begin{eqnarray}
0,1,2,3,\cdots, 2^{n}-1 \nonumber
 \end{eqnarray}     
can be written as 
   \begin{eqnarray} 
 k&=&a_{n-1}2^{n-1}+a_{n-2}2^{n-2}+\cdots+ a_{1}2^{1}+a_{0}2^{0}, \nonumber \\& & a_{n-1},\cdots, a_{0}=0,1.  \nonumber
    \end{eqnarray}     
  Then, the states  \eqref{base1}-\eqref{basen} can be expressed as 
  \begin{eqnarray}          
 \ket{k}=\ket{a_{n-1}a_{n-2}\cdots a_{1}a_{0}},  \nonumber
  \end{eqnarray}     
specifically
  \begin{eqnarray}   
          \ket{0}&=&\underbrace{\ket{00\cdots 0 0}}_{n},  \nonumber\\ \ket{1}&=&\underbrace{\ket{00\cdots 0 1}}_{n},  \nonumber\\ & \vdots&  \nonumber \\ \ket{2^{n}-1}&=&\underbrace{\ket{11\cdots 1 1}}_{n}.  \nonumber
 \end{eqnarray}
 Then, a state of a system with $n$ qubits can be written as
 \begin{eqnarray}   
 \ket{\psi}=\sum_{k=0}^{2^{n}-1}\alpha_{k}\ket{k}, \qquad    
\sum_{k=0}^{2^{n}-1}|\alpha_{k}|^{2}=1.  \nonumber
  \end{eqnarray}

  Now, the  QFT of a state $\ket{\psi}$ can be defined as 
 \begin{eqnarray} 
 {\rm QFT}\ket{\psi}&=\sum_{k=0}^{2^{n}-1}\tilde \alpha_{k}\ket{k}, \label{dqft}
  \end{eqnarray}
  where 
  \begin{eqnarray}
\tilde \alpha_{k}=\frac{1}{\sqrt{2^{n}}}  \sum_{j=0}^{2^{n}-1}\alpha_{j}\omega ^{kj} \label{cqft}
 \end{eqnarray}
 and 
   \begin{eqnarray}
         \omega=e^{\frac{\pi i}{2^{n-1}}}.  \nonumber
 \end{eqnarray}
Now, due  that QFT is a linear transformation, the following equation 
 \begin{eqnarray} 
 {\rm QFT}\ket{\psi}&=\sum_{j=0}^{2^{n}-1} \alpha_{j}{\rm QFT}\ket{j} \label{sbsqft}
  \end{eqnarray}
is satified.\\
 
In addition, notice that from the equations \eqref{dqft} and \eqref{cqft} we obtain
  \begin{eqnarray}
  {\rm QFT}\ket{\psi}&=&\sum_{k=0}^{2^{n}-1}\tilde \alpha_{k}\ket{k}= \sum_{k=0}^{2^{n}-1}\left( \frac{1}{\sqrt{2^{n}}} \sum_{j=0}^{2^{n}-1}\alpha_{j}\omega^{kj}\right)\ket{k }  \nonumber\\
  & =& \sum_{k=0}^{2^{n}-1}\sum_{j=0}^{2^{n}-1} \frac{1}{\sqrt{2^{n}}} \alpha_{j}\omega^{kj}\ket{k} =\sum_{j=0}^{2^{n}-1}\alpha_{j}\left(\frac{1}{\sqrt{2^{n}}} \sum_{k=0}^{2^{n}-1}\omega^{kj}\ket{k}\right), \nonumber
  \end{eqnarray}
 that is to say
 \begin{eqnarray}
{\rm QFT}\ket{\psi}=\sum_{j=0}^{2^{n}-1}\alpha_{j}\left( \frac{1}{\sqrt{2^{n}}} \sum_{k=0}^{2^{n}-1}\omega^{kj}\ket{k}\right).  \nonumber
 \end{eqnarray}
 Consequently, by using the equation  \eqref{sbsqft} we arrive at
 \begin{eqnarray}
 {\rm QFT}\ket{j}=\frac{1}{\sqrt{2^{n}}} \sum_{k=0}^{2^{n}-1}\omega^{kj}\ket{k}.  \nonumber
  \end{eqnarray}

\section{Alternative factorization}

Now, we can see  that if $n=1,$ we have   
 \begin{eqnarray}
  {\rm QFT}\ket{j}&=& \frac{1}{\sqrt{2}}\sum_{k=0}^{2^{1}-1}\omega^{kj}\ket{k} \nonumber\\&=& \frac{1}{\sqrt{2 }}\left(\omega^{0}\ket{0}+\omega^{j}\ket{1}\right),  \nonumber
 \end{eqnarray}
namely
 \begin{eqnarray}
 {\rm QFT}\ket{j}&=&\frac{1}{\sqrt{2 }} \left(\ket{0}+\omega^{j}\ket{1}\right), \quad  j=0,1. \label{1qft}
  \end{eqnarray}

Moreover, if   $n=2,$ the QFT is  
  \begin{eqnarray}
{\rm QFT}\ket{j}&=&\frac{1}{\sqrt{2^{2}}}\sum_{k=0}^{2^{2}-1}\omega^{kj}\ket{k}  \nonumber\\
&=&\frac{1}{\sqrt{2^{2}}}
\left(\omega^{0}\ket{0}+\omega^{j}\ket{1}+\omega^{2j}\ket{2}+\omega^{3j}\ket{3}\right)  \nonumber
\\&=&\frac{1}{\sqrt{2^{2}}} \left(\ket{00}+\omega^{j}\ket{01}+\omega^{2j}\ket{10}+\omega^{3j}\ket{11}\right), \nonumber
 \end{eqnarray}
that is to say
  \begin{eqnarray}
{\rm QFT}\ket{j}&=&\frac{1}{\sqrt{2^{2}}} \left(\ket{00}+\omega^{j}\ket{01}+\omega^{2j}\ket{10}+\omega^{3j}\ket{11}\right). \label{2qfac}
 \end{eqnarray}
This last state can be  separated in two states, one of this states is the QFT for $n=1.$ In fact,
 \begin{eqnarray}
  {\rm QFT}\ket{j} &=&\frac{1}{\sqrt{2^{2}}}
  \left(\ket{0}\ket{0}+\omega^{j}\ket{0}\ket{1}+\omega^{2j}\ket{1}\ket{0}+\omega^{2j}\omega^{j}\ket{1}\ket{1}\right)\nonumber\\
  &=&\frac{1}{\sqrt{2^{2}}} \left(\ket{0}\underbrace{\left(\ket{0}+\omega^{j}\ket{1}\right)}+\omega^{2j}\ket{1}\underbrace{\left(\ket{0}+\omega^{j}\ket{1} \right)} \right) \nonumber \\
  &=&\frac{1}{\sqrt{2^{2}}} \left(\ket{0}+\omega^{2j}\ket{1}\right)\underbrace{ \left(  \ket{0}+\omega^{j}\ket{1} \right) }_{{\rm QFT} \,\, {\rm for} \,\, n=1}.  \nonumber
\end{eqnarray}
 Then, we obtain
 \begin{eqnarray}            
  {\rm QFT}\ket{j}=\frac{1}{\sqrt{2^{2}}} \left(\ket{0}+\omega^{2j}\ket{1}\right)\left(\ket{0}+\omega^{j}\ket{1}\right),\quad j=0,1,2,3. \label{2qft}
  \end{eqnarray}

Consequently,  if the QFT  for   the case $n=1$  \eqref{1qft} is multiplied on the left for  the state  
 \begin{eqnarray}
\frac{1}{\sqrt{2 } }\left(\ket{0}+\omega^{2j}\ket{1}\right),  \nonumber
  \end{eqnarray}        
 the  QFT for case $n=2$  \eqref{2qft} is gotten. \\

Furthermore, when $n=3,$ we obtain      
 \begin{eqnarray}            
{\rm QFT}\ket{j}&=&\frac{1}{\sqrt{2^{3}}}  \sum_{k=0}^{2^{3}-1}\omega^{kj}\ket{k} \nonumber \\&=&\frac{1}{\sqrt{2^{3}}} \bigg(\omega^{0}\ket{0}+\omega^{j}\ket{1}+\omega^{2j}\ket{2}+\omega^{3j}\ket{3}+\nonumber \\
 & &+\omega^{4j}\ket{4}+\omega^{5j}\ket{5}+\omega^{6j}\ket{6}+\omega^{7j}\ket{7}\bigg) \nonumber \\ &=& \frac{1}{\sqrt{2^{3}}} \bigg(\ket{000}+\omega^{j}\ket{001}+\omega^{2j}\ket{010}+\omega^{3j}\ket{011}+\nonumber\\ & &+\omega^{4j}\ket{100}+\omega^{5j}\ket{101}+\omega^{6j}\ket{110}+\omega^{7j}\ket{111}\bigg), \nonumber
 \end{eqnarray}
 in other words 
 \begin{eqnarray}            
{\rm QFT}\ket{j}&=& \frac{1}{\sqrt{2^{3}}} \bigg(\ket{000}+\omega^{j}\ket{001}+\omega^{2j}\ket{010}+\omega^{3j}\ket{011}+ \nonumber  \\ & &+\omega^{4j}\ket{100}+\omega^{5j}\ket{101}+\omega^{6j}\ket{110}+\omega^{7j}\ket{111}\bigg). \nonumber
 \label{preqft2}
 \end{eqnarray}
This last state can be  separated in two states, one of this states is the QFT for $n=2.$ Actually,
 \begin{eqnarray}           
{\rm QFT}\ket{j}&=&\frac{1}{\sqrt{2^{3}}}  \bigg(\ket{0}\ket{00}+\omega^{j}\ket{0}\ket{01}+\omega^{2j}\ket{0}\ket{10}+\omega^{3j}\ket{0}\ket{11}+\nonumber\\
& &+\omega^{4j}\ket{1}\ket{00}+\omega^{4j}\omega^{j}\ket{1}\ket{01}+\omega^{4j}\omega^{2j}\ket{1}\ket{10}+\omega^{4j}\omega^{3j}\ket{1}\ket{11} \bigg)\nonumber \\
&=& \frac{1}{\sqrt{2^{3}}} \bigg(\ket{0}\left(\ket{00}+\omega^{j}\ket{01}+\omega^{2j}\ket{10}+\omega^{3j}\ket{11}\right)+\nonumber\\
& &+\omega^{4j}\ket{1}\left(\ket{00}+\omega^{j}\ket{01}+\omega^{2j}\ket{10}+\omega^{3j}\ket{11}\right) \bigg)\nonumber\\
&=& \frac{1}{\sqrt{2^{3}}} \left(\ket{0}+\omega^{4j}\ket{1}\right)\left(\ket{00}+\omega^{j}\ket{01}+\omega^{2j}\ket{10}+\omega^{3j}\ket{11}\right). \nonumber
 \end{eqnarray}
that is       
 \begin{eqnarray}            
{\rm QFT}\ket{j}= \frac{1}{\sqrt{2^{3}}} \left(\ket{0}+\omega^{4j}\ket{1}\right)\underbrace{\left(\ket{00}+\omega^{j}\ket{01}+\omega^{2j}\ket{10}+\omega^{3j}\ket{11} \right) }_{{\rm QFT} \,\, {\rm for} \,\, n=2}. \nonumber
 \end{eqnarray}
Then, by using the  equations \eqref{preqft2} and  \eqref{2qft} we arrive at
 \begin{eqnarray}        
 {\rm QFT}\ket{j}&=& \frac{1}{\sqrt{2^{3}}} \left(\ket{0}+\omega^{4j}\ket{1}\right)\left(\ket{0}+\omega^{2j}\ket{1}\right)\left(\ket{0}+\omega^{j}\ket{1}\right), \label{3qft}\\
& & j=0,1,2,3,4,5,6,7.  \nonumber
 \end{eqnarray}

Hence, if the QFT  for the  case $n=2$  \eqref{2qft} is multiplied on the left for  the state  
 \begin{eqnarray}
\frac{1}{\sqrt{2 } }\left(\ket{0}+\omega^{4j}\ket{1}\right),  \nonumber
  \end{eqnarray}        
 the  QFT for case $n=3$  \eqref{3qft} is obtained. \\

In general, we can see that  if the state ${\rm QFT}\ket{j}$ of $n-1$ qubits is multiplied d on the left  for 
 \begin{eqnarray}
\frac{1}{\sqrt{2 }}\left(\ket{0}+\omega^{2^{n-1}j}\ket{1}\right)  \nonumber
 \end{eqnarray} 
the state ${\rm QFT}\ket{j}$ for  $n$ qubits is obtained. \\
 
For example, if $n=4,$ by using the equation \refeq{3qft} and the state
 \begin{eqnarray}
\frac{1}{\sqrt{2 }}\left(\ket{0}+\omega^{2^{4-1}j}\ket{1}\right)  \nonumber
 \end{eqnarray} 
we have
 \begin{eqnarray}
 {\rm QFT}\ket{j}&=&  \frac{1}{\sqrt{2^{4}}}\left(\ket{0}+\omega^{8j}\ket{1}\right)\left(\ket{0}+\omega^{4j}\ket{1}\right)\left(\ket{0}+\omega^{2j}\ket{1}\right)\left(\ket{0}+\omega^{j}\ket{1}\right), \nonumber \\
 &&j=0,1,2,3,\cdots,15. \nonumber
  \end{eqnarray}
 In general we have     
 \begin{eqnarray}            
 {\rm QFT}\ket{j}&=& \frac{1}{\sqrt{2^{n} }} \left(\ket{0}+\omega^{2^{n-1}j}\ket{1}\right)\left(\ket{0}+\omega^{2^{n-2}j}\ket{1}\right)\cdots \left(\ket{0}+\omega^{2j}\ket{1}\right)\left(\ket{0}+\omega^{j}\ket{1}\right),\qquad   \label{qft}\\ & &j=0,1,2,3,\cdots,2^{n}-1. \nonumber
  \end{eqnarray}

In the next section, we proof that this factorization  allows to built  an alternative quantum circuit for the QFT, which is faster than the usual. \\

In addition, if 
   \begin{eqnarray} 
 j&=&b_{n-1}2^{n-1}+b_{n-2}2^{n-2}+\cdots+ b_{1}2^{1}+b_{0}2^{0}, \nonumber \\& & b_{n-1},\cdots, b_{0}=0,1, \nonumber
    \end{eqnarray}     
  by using the fractional binary notation, 
 \begin{eqnarray}
  [0.x_{1}x_{2}\cdots x_{n}]=\frac{x_{1}}{2}+ \frac{x_{2}}{2^{2}}+ \frac{x_{3}}{2^{3}}+ \cdots+\frac{x_{n}}{2^{n}},  \nonumber 
   \end{eqnarray}
the QFT  \eqref{qft} can be written as 
 \begin{eqnarray}
 {\rm QFT} \ket{j}&=&{\rm QFT} \ket{b_{n-1}\cdots b_{1}b_{0}} \nonumber\\
 &=& \frac{1}{\sqrt{2^{n} }} \left(\ket{0} +e^{2\pi i[0.b_{0}]} \ket{1}\right)\left( \ket{0} + e^{2\pi i[0.b_{1}b_{0}]}\ket{1} \right) \cdots \left( \ket{0} + e^{2\pi i[0.b_{n}\cdots b_{1}b_{0}]} \ket{1} \right),\nonumber
 \end{eqnarray}
 which is the usual expression of the QFT.

\section{Quantum circuit}
\label{AQC}
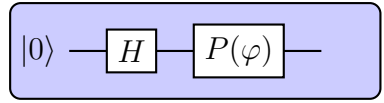
\begin{figure}[hbt!]
\centering
\begin{quantikz}
\gategroup[wires=1,steps=6,style={rounded corners,fill=blue!20}, background]{}
&\lstick{$|{0}\rangle$} & \gate{H}& \gate{P(\varphi)}&  \qw & 
\end{quantikz}
\caption{Quantum circuit for the state \eqref{sbb}}
    \label{fig:sbb}
\end{figure}

Notice that the states
  \begin{eqnarray}           
\frac{1}{\sqrt{2}}\left(\ket{0}+e^{i\varphi}\ket{1}\right)  \label{bbs}
 \end{eqnarray}       
are the building blocks of the QFT \eqref{qft}. The states \eqref{bbs} can be constructed with the Hadamard gate
  \begin{eqnarray}           
H=\frac{1}{\sqrt{2}}\begin{pmatrix}1 & 1\\ 1 & -1 \end{pmatrix}   \nonumber
 \end{eqnarray}       
and the  phase shift gate  
  \begin{eqnarray}           
  P(\varphi)=\begin{pmatrix}1 & 0\\ 0& e^{i\varphi} \end{pmatrix} .  \nonumber
 \end{eqnarray}       
Certainly, we have 
  \begin{eqnarray}           
P(\varphi) H\ket{0}=\frac{1}{\sqrt{2}}\left(\ket{0}+e^{i\varphi}\ket{1}\right). \label{sbb}
 \end{eqnarray}       
The quantum circuit  for this last expression is given in  the Figure \ref{fig:sbb}.

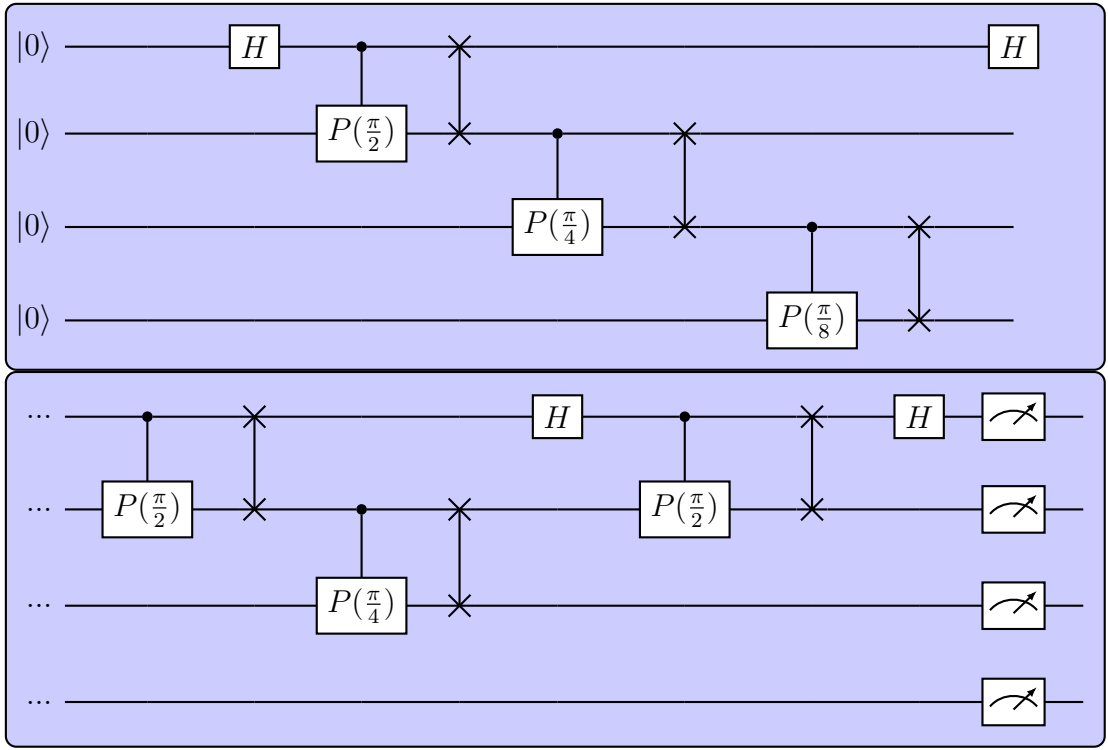
\begin{figure}[H]
\centering
\begin{quantikz}
\gategroup[wires=4,steps=12,style={rounded corners,fill=blue!20}, background]{}
&\lstick{$|{0}\rangle$} & \qw & \gate{H} & \ctrl{1} & \swap{1} & \qw      & \qw      & \qw      & \qw & \gate{H} \\
&\lstick{$|{0}\rangle$} & \qw & \qw      & \gate{P(\frac{\pi}{2})} & \swap{0}  & \ctrl{1} & \swap{1} & \qw      & \qw & \qw   \\
&\lstick{$|{0}\rangle$} & \qw & \qw      & \qw      & \qw      & \gate{P(\frac{\pi}{4})} & \swap{0}  & \ctrl{1} & \swap{1} & \qw    \\
&\lstick{$|{0}\rangle$} & \qw & \qw      & \qw      & \qw      & \qw                     & \qw      & \gate{P(\frac{\pi}{8})} & \swap{0}  & \qw    \\
\gategroup[wires=4,steps=12,style={rounded corners,fill=blue!20}, background]{}
&\lstick{...} & \ctrl{1} & \swap{1} & \qw      & \qw & \gate{H} & \ctrl{1} & \swap{1} & \gate{H} & \meter{} & \qw \\
&\lstick{...} & \gate{P(\frac{\pi}{2})} & \swap{0}  & \ctrl{1} & \swap{1} & \qw & \gate{P(\frac{\pi}{2})} & \swap{0} & \qw & \meter{} & \qw  \\
&\lstick{...} & \qw      & \qw      & \gate{P(\frac{\pi}{4})} & \swap{0}  & \qw & \qw & \qw & \qw & \meter{} & \qw  \\
&\lstick{...} & \qw      & \qw      & \qw                     & \qw      & \qw & \qw & \qw & \qw & \meter{} & \qw
\end{quantikz}

\caption{Quantum circuit for the QFT \eqref{qft}}
    \label{fig:qft}
\end{figure}

The novel quantum circuit for the complete  QFT is given in the Figure   \ref{fig:qft} and the Qiskit code for this circuit is given in the Listing \ref{lst:qft}.
\begin{lstlisting}[language=Python, caption=Qiskit code to obtain novel the  QFT  \eqref{qft},  label={lst:qft} ]
import qiskit as q
from qiskit import QuantumCircuit
from qiskit.quantum_info import Statevector
from numpy import pi
import numpy as np
import time

def cz_pow(qc, control, target, power):
    qc.cp(pi * power, control, target)

def cz_and_swap(qc, q0, q1, rot):
    cz_pow(qc, q0, q1, rot)
    qc.swap(q0, q1)

qc = QuantumCircuit(4, 4)

a, b, c, d = 0, 1, 2, 3

qc.h(range(4))
qc.h(a)
cz_and_swap(qc, a, b, 0.5)
cz_and_swap(qc, b, c, 0.25)
cz_and_swap(qc, c, d, 0.125)
qc.h(a)
cz_and_swap(qc, a, b, 0.5)
cz_and_swap(qc, b, c, 0.25)
qc.h(a)
cz_and_swap(qc, a, b, 0.5)
qc.h(a)
qc.measure(range(4), range(4))
simulator = AerSimulator()
qc_t = transpile(qc, simulator, optimization_level=0)
result  = simulator.run(qc_t, shots=1024).result()
counts  = result.get_counts()
\end{lstlisting}

The new code given in the Listing \ref{lst:qft}  improves the standard one over $0.0004$ seconds.

\section{Alternative HHL algorithm  } 
\label{AHHL}

The HHL algorithm is a quantum method  for solving linear systems of equations \cite{HHL} . 
 By using the novel quantum circuit  \ref{fig:qft},  we propose an alternative code for the HHL algorithm.\\
 
 The Qiskit code for this alternative HHL algorithm using 15 qubits is given in the Listing. \ref{lst:AHHL}.

\begin{lstlisting}[language=Python, caption=Qiskit code for alternative HHL using 15 qubits. ,  label={lst:AHHL} ]
import qiskit as qimport numpy as np
from qiskit import QuantumRegister, ClassicalRegister, QuantumCircuit, transpile
from qiskit_aer import AerSimulator
from qiskit.visualization import plot_histogram
import matplotlib.pyplot as plt
import time

num_qpe_qubits = 6
num_solution_qubits = 8
num_auxiliary_qubits = 1

total_qubits = num_qpe_qubits + num_solution_qubits + num_auxiliary_qubits



def iqft_manual_general(qc_arg, q_reg, num_qubits):
    for i in range(num_qubits // 2):
        qc_arg.swap(q_reg[i], q_reg[num_qubits - 1 - i])

  
    for i in reversed(range(num_qubits)):
        qc_arg.h(q_reg[i])
        for j in reversed(range(i)):
            qc_arg.cp(-np.pi / (2**(i - j)), q_reg[j], q_reg[i])



qr = QuantumRegister(total_qubits, 'q')
cr = ClassicalRegister(num_solution_qubits + num_auxiliary_qubits, 'c')
qc = QuantumCircuit(qr, cr)


for i in range(num_qpe_qubits, num_qpe_qubits + num_solution_qubits):
    qc.h(qr[i])
qc.barrier()


for i in range(num_qpe_qubits):
    qc.h(qr[i])
    for j in range(num_solution_qubits):
        qc.cu(0, 0, np.pi / (2**(i + j + 1)), 0, qr[i], qr[num_qpe_qubits + j])
qc.barrier()


iqft_manual_general(qc, qr[0:num_qpe_qubits], num_qpe_qubits)
qc.barrier()


for i in range(num_qpe_qubits):
    qc.cry(np.pi / (2**(i + 1)), qr[i], qr[total_qubits - 1])
qc.barrier()


qc.measure(qr[num_qpe_qubits : num_qpe_qubits + num_solution_qubits],
           cr[0 : num_solution_qubits])
qc.measure(qr[total_qubits - 1], cr[num_solution_qubits])


print(qc)

simulator = AerSimulator()
qc_t = transpile(qc, simulator, optimization_level=0)


result = simulator.run(qc_t, shots=1024).result()
counts = result.get_counts()

plot_histogram(counts)

plt.show()
\end{lstlisting}

The novel code for the alternative HHL algorithm  improves the usual one for $0.0061$ seconds.

\section{Modified  Shor's algorithm   } 
\label{Shor}

Shor’s algorithm is efficient for factoring integers and is  one of most relevant algorithms in quantum cryptography  \cite{Shor}. \\

 By using the novel quantum circuit  \ref{fig:qft},   we propose an modified  code for the Shor’s algorithm \\

 The Qiskit code for this alternative Shor's algorithm  using 4 qubits is given in the Listing. \ref{lst:Shor}.

\begin{lstlisting}[language=Python, caption=Qiskit code for alternative Shor using 4 qubits. ,  label={lst:Shor} ]


from fractions import Fraction
from math import gcd, pi
from qiskit import QuantumCircuit, transpile
from qiskit_aer import AerSimulator
from qiskit.visualization import plot_histogram
import matplotlib.pyplot as plt
import time

N = 15
a = 7
n_input = 4
n_output = 4

qc = QuantumCircuit(n_input + n_output, n_input)

qc.h(range(n_input))

qc.x(n_input)
qc.barrier()

def c_amod15(circuit, control_qubit):
    circuit.cswap(control_qubit, n_input + 0, n_input + 2)
    circuit.cswap(control_qubit, n_input + 1, n_input + 3)
    circuit.cswap(control_qubit, n_input + 0, n_input + 1)
    circuit.cswap(control_qubit, n_input + 1, n_input + 2)
    circuit.cswap(control_qubit, n_input + 2, n_input + 3)

for i in range(n_input):
    if i == 0: c_amod15(qc, i)
    if i == 1: c_amod15(qc, i); c_amod15(qc, i)

qc.barrier()

def iqft(circuit, q0, q1, q2, q3):
    circuit.h(q0)
    circuit.swap(q0, q1); circuit.cp(-pi * 0.5, q0, q1)
    circuit.h(q0)
    circuit.swap(q1, q2); circuit.cp(-pi * 0.25, q1, q2)
    circuit.swap(q0, q1); circuit.cp(-pi * 0.5, q0, q1)
    circuit.h(q0)
    circuit.swap(q2, q3); circuit.cp(-pi * 0.125, q2, q3)
    circuit.swap(q1, q2); circuit.cp(-pi * 0.25, q1, q2)
    circuit.swap(q0, q1); circuit.cp(-pi * 0.5, q0, q1)
    circuit.h(q0)


iqft(qc, 0, 1, 2, 3)
qc.barrier()

qc.measure(range(n_input), range(n_input))

simulator = AerSimulator()
qc_t = transpile(qc, simulator, optimization_level=0)
start_time = time.time()
result = simulator.run(qc_t, shots=2048).result()
end_time = time.time()
counts = result.get_counts()
\end{lstlisting}

The novel code for the alternative Shor's algorithm improves the usual one for $0.0061$ seconds.

\section{Conclusions}
\label{Con}

The Quantum Fourier Transform (QFT) is one of the most fundamental algorithms in quantum computing. In this paper, we presented an alternative method for factoring the QFT. Inspired by this approach, we introduced a novel quantum circuit for implementing the QFT. Furthermore, we demonstrated that the new algorithm is faster than the standard one. As an application of this new circuit, we proposed alternative versions of the HHL algorithm and Shor's algorithm, demonstrating better performance than their usual implementations.\\

In future works, we will study additional applications of this novel quantum circuit in areas such as 
quantum machine learning,  physics,  quantum chemistry, finance, etc.

\section*{Acknowledgements}
 E. M-G,  G. C, and R.C. R. are  supported by CONAHCyT Master fellowship.

\section*{Authors contributions}
All authors contributed equally to this work.

\section*{Conflict of interest}
Authors declare that they have no conflict of
interest.

\section*{Ethical approval}
This article does not contain any studies with
human participants or animals performed by any of the authors.

\section*{Declaration of competing interest}
The authors declare that they have no known competing financial
interests or personal relationships that could have appeared to influence
the work reported in this paper.

\section*{Data availability}
No data was used for the research described in the article.

\end{document}